\documentclass[aps,prl,twocolumn,superscriptaddress]{revtex4-1}
\setcitestyle{authoryear,round}
\usepackage{graphicx}
\usepackage{amsmath}
\usepackage{bm}
\usepackage[version=4]{mhchem}
\usepackage{upgreek}
\usepackage{hyperref}
\usepackage{float}
\usepackage{siunitx}
\usepackage{capt-of}
\usepackage[dvipsnames]{xcolor}
\usepackage{tabularx}
\usepackage{booktabs}
\usepackage{type1cm}
\usepackage{natbib}
\usepackage{caption}
\usepackage{array}

\begin{document}

\title{Graph Neural Network Prediction of Infrared Spectra of Interstellar Polycyclic Aromatic Hydrocarbons}

\author{Guoqing Tang}
\affiliation{Laboratory for Relativistic Astrophysics, Department of Physics, Guangxi University, 530004 Nanning, China}
\affiliation{School of Computer Science and Technology, Chongqing University of Posts and Telecommunications, Chongqing 400065,China}

\author{Jiang He}
\affiliation{Laboratory for Relativistic Astrophysics, Department of Physics, Guangxi University, 530004 Nanning, China}

\author{Zhao Wang}
\email{zw@gxu.edu.cn}
\affiliation{Center for Applied Mathematics of Guangxi (Guangxi University), Nanning 530004, China}
\affiliation{Laboratory for Relativistic Astrophysics, Department of Physics, Guangxi University, Nanning 530004, China}

\author{Dong Qiu}
\affiliation{Center for Applied Mathematics of Guangxi (Guangxi University), Nanning 530004, China}
\affiliation{School of Mathematics and Information Science, Guangxi University, Nanning 530004, China}

\begin{abstract}

Polycyclic aromatic hydrocarbons (PAHs) are recognized as the primary contributors to the aromatic infrared bands (AIBs) widely observed in space. However, analyzing these AIBs remains challenging because of the immense structural diversity within the PAH family, which makes the computation of reliable reference spectra difficult. To address this, we developed an efficient graph neural network (GNN) framework that can predict PAH absorption spectra up to 10,000 times faster than traditional quantum chemical methods. We evaluated four representative GNN architectures, including graph convolutional network (GCN), graph attention network (GAT), message passing neural network (MPNN), and attentive fingerprint (AFP). The AFP model is found to deliver the best overall performance and is further trained using five different spectral distance metrics as loss functions, among which the Jensen-Shannon divergence yields the most accurate and stable results. The model performs best for PAHs containing 20-40 carbon atoms, while accuracy decreases for larger molecules, reflecting the limited availability of training data. Overall, this framework offers a fast method to generate approximate reference spectra for small- to medium-sized PAHs, supporting future AIB analysis.

\end{abstract}


\maketitle

\section{Introduction}
Polycyclic aromatic hydrocarbons (PAHs), molecules composed of fused aromatic rings, play a central role in astrochemistry as proposed carriers of the unidentified infrared emission (UIE) features observed throughout interstellar and circumstellar environments \citep{Tielens2008}. These prominent mid-infrared bands, typically appearing at 3.3, 6.2, 7.7, 8.6, 11.2, and \SI{12.7}{\micro\meter}, arise from \ce{C-H} and \ce{C-C} vibrational modes in PAH molecules \citep{Peeters2011}. Since their initial identification as UIE carriers in the 1980s \citep{leger1984identification, allamandola1985polycyclic}, substantial efforts have been devoted to characterizing PAH populations in terms of size, structure, charge state, and abundance, providing critical insights into the physical conditions of star-forming regions, photodissociation zones, and stellar outflows \citep{Allamandola1989, hrodmarsson2025astropah}.

Despite the widespread detection of these PAH-related IR features, identifying specific PAH molecules in space remains challenging, largely due to their immense structural diversity and the complex relationship between molecular structure and IR signatures \citep{li2020spitzer,Hanine2020PAHFormation, Qi2018AdsorptionOLC}. Although a few PAHs have recently been identified via high-resolution radio observations \citep{mcguire2018detection, mcguire2021detection, burkhardt2021discovery}, a comprehensive understanding of the molecular origins of UIE features and their band correlations is still lacking. PAH IR spectra are highly sensitive to subtle molecular characteristics, including edge configuration, charge state, heteroatom substitution, functional groups, size, and symmetry \citep{peeters2021spectroscopic}, highlighting the need for robust computational models that can reliably link molecular structure to vibrational spectra \citep{peeters2021spectroscopic, Li2024Cyanonaphthalenes,Li2024Indene}.

Quantum chemical calculations (QCCs) remain the gold standard for computing molecular IR spectra, yet their high computational cost severely limits large-scale exploration of interstellar PAH chemical space. While extensive density functional theory (DFT) spectral libraries such as the NASA Ames PAH database (PAHdb) \citep{bauschlicher2010nasa, boersma2014nasa, bauschlicher2018nasa} have enabled systematic studies, their limited molecular coverage motivates the use of machine learning (ML) models to interpolate within the vast PAH chemical space. ML has emerged as a promising alternative, with early feedforward neural networks demonstrating that structure–spectrum relationships can be learned from molecular fingerprints at speeds orders of magnitude faster than QCCs \citep{kovacs2020machine}. Subsequent applications of random forest models and other data-driven approaches have further confirmed ML's ability to capture correlations between molecular structure and spectral features \citep{meng2023machine, calvo2021infrared, liu2024infrared, zapata2021computational}. However, these methods often treat molecules as flat feature vectors, overlooking their intrinsic graph-like structure.

Deep learning architectures such as graph neural networks (GNNs) directly address this limitation by representing atoms as nodes and bonds as edges, enabling hierarchical, context-aware molecular representations through iterative message-passing \citep{mcgill2021predicting, Saquer2024Infrared, Stienstra2024GraphormerIR, ShabanTameh2024IRPrediction, xu2025pretrained, singh2022graph, stienstra2025machine, beglaryan2025toward}. This architectural advantage makes GNNs particularly well-suited for predicting vibrational spectra. Recent advances in equivariant neural networks and transformer-based spectral models have demonstrated significantly improved performance in molecular spectra prediction \citep{Nequip, Allegro, TorchMD-NET, PaiNN, FCNN-Spec}. However, most of these approaches rely on explicit geometric or physics-constrained features, highlighting an important limitation of purely SMILES-based GNN models.

In this work, we present a GNN-based framework for predicting PAH IR spectra using data from PAHdb. We systematically examine how different spectral distance metrics, employed as loss functions across various GNN architectures, influence predictive performance. Our goal is to develop models that provide both accurate and chemically interpretable representations of PAH IR spectral features, thereby facilitating their identification in astronomical environments.

\section{Methods}
\subsection{GNN Architectures}

GNNs provide a powerful framework for molecular property prediction by representing molecules as graphs with atoms as nodes and chemical bonds as edges. These models operate through iterative information propagation and aggregation across molecular graphs, where at each layer, atom-level features are updated by combining their current representations with those of neighboring atoms via message passing along chemical bonds. This hierarchical process enables the networks to capture both local atomic environments and global topological patterns, ultimately generating comprehensive molecular representations.

This architecture is particularly well-suited for predicting PAH spectra, as their spectral characteristics are strongly influenced by conjugation effects, aromatic ring connectivity, and subtle variations in local bonding environments. While classical GNNs such as Graph Attention Network (GAT), Message Passing Neural Network (MPNN), and Graph Convolutional Network (GCN) operate primarily on the message-passing paradigm, recent architectures have extended beyond this framework. Attentive Fingerprint (AFP), for instance, employs a global attention mechanism to capture long-range dependencies within molecular structures \citep{xiong2019attentivefp}.

This study evaluates these four distinct GNN architectures, AFP, GAT, MPNN, and GCN, to assess their performance in predicting PAH infrared (IR) spectra. For comparative analysis, we also include a classical Molecular Fingerprint (MFP) approach as a non-GNN baseline, providing a reference point against the graph-based methods. We note that the GNNs used here operate solely on topological graphs derived from SMILES and do not incorporate 3D geometries or vibrational features, which may fundamentally limit their ability to generalize to structurally diverse PAHs.

For spectral prediction tasks, defining appropriate distance metrics is essential for quantifying discrepancies between predicted and reference spectra. We investigate five specialized metrics that capture different aspects of spectral similarity: Earth Mover's Distance (EMD), Jensen–Shannon Divergence (JSD), Hellinger Distance (HD), Total Variation Distance (TVD), and Spectrum Information Similarity (SIS, \citet{mcgill2021predicting}). This multi-metric approach enables a comprehensive assessment of prediction accuracy across various spectral characteristics.

\begin{figure*}
\centering
\includegraphics[width=0.7\textwidth]{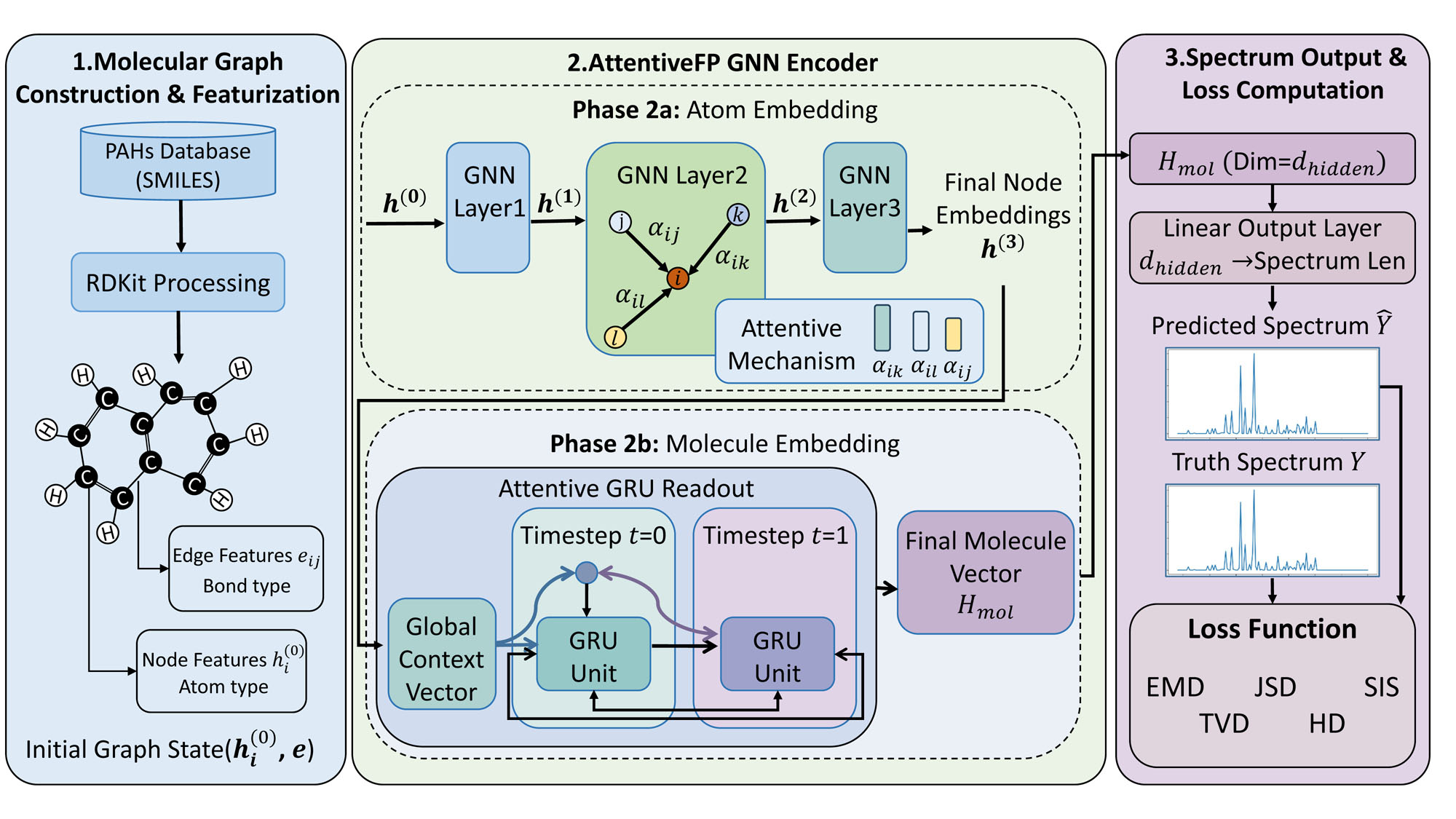}
\caption{Workflow for training an AttentiveFP for IR spectrum computation. SMILES strings from the PAHdb are converted by RDKit into molecular graphs with atom (node) and bond (edge) features. An AttentiveFP encoder performs attention-based message passing over three GNN layers to obtain atom embeddings, followed by an iterative GRU readout that aggregates them into a molecule-level embedding \(H_{\mathrm{mol}}\). A linear output layer maps \(H_{\mathrm{mol}}\) to the infrared spectrum \(\hat{Y}\), and the prediction is trained against the reference spectrum \(Y\) using EMD as the primary loss, with JSD, SIS, TVD, and HD available as alternatives.}
\label{F1}
\end{figure*}

Our computational pipeline comprises four sequential steps, as illustrated in Figure \ref{F1}:

\begin{enumerate}
\item \textbf{Molecular Representation and Data Preparation}: Atomic coordinates are converted into Simplified Molecular Input Line Entry System (SMILES) string representations. These unstructured inputs (molecule SMILES + raw spectrum) are then transformed into structured graph data compatible with GNN processing.

\item \textbf{Architecture Selection}: We compare four GNN-based models, AFP, GCN, GAT, and MPNN, against the MFP baseline. Based on comprehensive performance evaluation, AFP is selected as the foundational architecture due to its superior predictive capability.

\item \textbf{Loss Function Optimization}: Using AFP as the base architecture, we systematically evaluate five spectral distance metrics (EMD, JSD, HD, TVD, SIS) as loss functions during training. Through multi-metric performance comparison, JSD emerges as the optimal choice, yielding the final best model (AFP+JSD).

\item \textbf{Spectral Prediction and Error Analysis}: The optimized AFP+JSD model is employed for spectral prediction across the dataset. We systematically evaluate the overall agreement between predicted and actual spectra, with particular emphasis on error distribution patterns across high- and low-frequency regions to assess model performance across different spectral domains.
\end{enumerate}

The source code and datasets used to train the GNN and predict IR spectra are publicly available, as noted in the Data Availability Statement, to facilitate study reproducibility.

\subsection{Dataset}
The dataset for this study was sourced from the computational datasets of PAHdb. The primary training and validation data comprise 1,570 neutral PAH molecules from version 3.2, each containing up to 47 C atoms. Most of the spectra were computed via density functional theory at the B3LYP level \citep{liao2023density, meng2023evolution}. To evaluate the model's generalizability to large molecules, we additionally employed 997 neutral PAH molecules from version 4.0$\alpha$, with C atom counts ranging from 50 to 100.

For each molecule, histogram representations were generated using a bin width of \SI{18.23}{\per\centi\meter} determined via Knuth's Bayesian binning rule \citep{knuth2006optimal}, which provides a coarse but computationally manageable discretization. We acknowledge that such binning may broaden narrow PAH vibrational features and can limit the model’s ability to resolve fine spectral structure. This yielded 309 bins spanning the frequency range from \SIrange{0.21}{5603.63}{\per\centi\meter}. Bins beyond the 308th bin (\SI{5585.49}{\per\centi\meter}) were excluded from analysis, as they contained only a single molecular entry and were considered potential outliers or noise artifacts.

\begin{table}[h]
\centering
\caption{Performance comparison of models using different frequency ranges and bin widths, evaluated using JSD as the error metric.}
\label{tab:jsd_comparison}
\setlength{\extrarowheight}{3pt} 
\begin{tabular}{lcc}
\hline\hline 
\textbf{Frequency Range} & \multicolumn{2}{c}{\textbf{Mean JSD Error}}\\
(cm$^{-1}$) & {high frequency} & {low frequency} \\
\hline
100.0 - 5603.63 & 0.030 & 0.102 \\
0.21 - 5603.63  & 0.029 & 0.097 \\
\hline
\textbf{Bin Width} (cm$^{-1}$) & & \\
\hline
5.0             & 0.121 & 0.308 \\
10.0            & 0.051 & 0.186 \\
18.23           & 0.029 & 0.097 \\
\hline\hline 
\end{tabular}
\end{table}

We performed a sensitivity test on the model's performance by using a lower frequency limit of ~\SI{100}{\per\centi\meter} \citep{Zhang2010FarIR} and narrower bin widths of \SI{5}{\per\centi\meter} and \SI{10}{\per\centi\meter}. The results are shown in Table~\ref{tab:jsd_comparison} using JSD as a performance metric. We find that changing the lower limit (which also alters the bin width) only causes a minor change in model performance, while using narrower bin widths reduces performance. This highlights an important principle in machine learning: optimal algorithmic settings are not always the same as those preferred for human interpretation. The model performs best with a slightly coarser binning, which likely helps it generalize patterns more effectively by reducing sensitivity to minor noise or variations in the data.

All spectra were normalized to their respective maximum intensities to focus on relative spectral patterns. We also evaluated global normalization (scaling all spectra by the dataset maximum intensity) but found it resulted in inferior model performance. Following normalization, each spectrum was partitioned into low-frequency (\SIrange{0.21}{2339.50}{\per\centi\meter}) and high-frequency (\SIrange{2339.50}{5585.49}{\per\centi\meter}) components at the 128th bin, corresponding to a natural gap in the spectral distribution. These segmented spectra served as separate target outputs during model training.

\subsection{Molecular Representation}
PAH molecular structures were initially represented as canonical SMILES strings. Rather than employing character-level tokenization, we directly parsed each SMILES string into a molecular object using the \texttt{RDKit} toolkit. This approach avoids potential ambiguities from multi-character tokens (e.g., ``Cl'' or ring indices) while preserving complete molecular topology.

The resulting molecular objects were transformed into graph-structured data compatible with GNNs, consisting of three primary components:

\begin{itemize}
\item \textbf{Atom feature matrix}: Each atom is represented by an $11$-dimensional one-hot vector encoding its element type (C, H, O, N, S, F, Cl, Br, I, P, or ``other''), supplemented by a scalar feature indicating atomic degree (i.e., number of covalent bonds). This yields a $12$-dimensional feature vector per atom.

\item \textbf{Edge feature matrix}: Chemical bonds are encoded as $4$-dimensional one-hot vectors representing bond types: single, double, triple, or aromatic. Aromatic bonds are particularly crucial for capturing the conjugated ring systems characteristic of PAHs.

\item \textbf{Edge index}: Molecular connectivity is encoded in a $2 \times E$ tensor (where $E$ denotes the total number of bonds). To enable bidirectional message passing in GNNs, each bond is represented by two directed edges ($i \to j$ and $j \to i$).
\end{itemize}

For the MFP baseline model, SMILES strings were converted into feature vectors using the CircularFingerprint tool from the DeepChem library. The normalized PAH IR spectrum was stored as the target label tensor, while original SMILES strings were retained for downstream analysis. This graph-based representation was subsequently used as input for GNN training and evaluation.

\subsection{Model training}
We evaluated five established models, AFP, GCN, GAT, MPNN, and MFP, each employing distinct information propagation and aggregation mechanisms. The GNN architectures introduce unique inductive biases: GCN uses layer-wise convolutions for neighborhood aggregation; GAT incorporates attention mechanisms to weight neighbor influences; AFP learns task-specific attention scores to emphasize chemically relevant substructures; and MPNN learns both message functions and update rules to model complex atomic interactions. The MFP baseline provides a fixed, interpretable representation for comparison.

For optimization, Earth Mover’s Distance (EMD) \citep{monge1781} was used as the primary loss function, chosen for its effectiveness in comparing spectral distributions. Four additional loss functions were evaluated to comprehensively assess model performance. The total loss was computed as the sum of separate low-frequency and high-frequency region losses. All predicted spectra were normalized relative to the maximum intensity of the combined true and predicted values within each sample group to ensure consistent evaluation.

\section{Results and Discussions}

\begin{figure}
\centering
\includegraphics[width=0.46\textwidth]{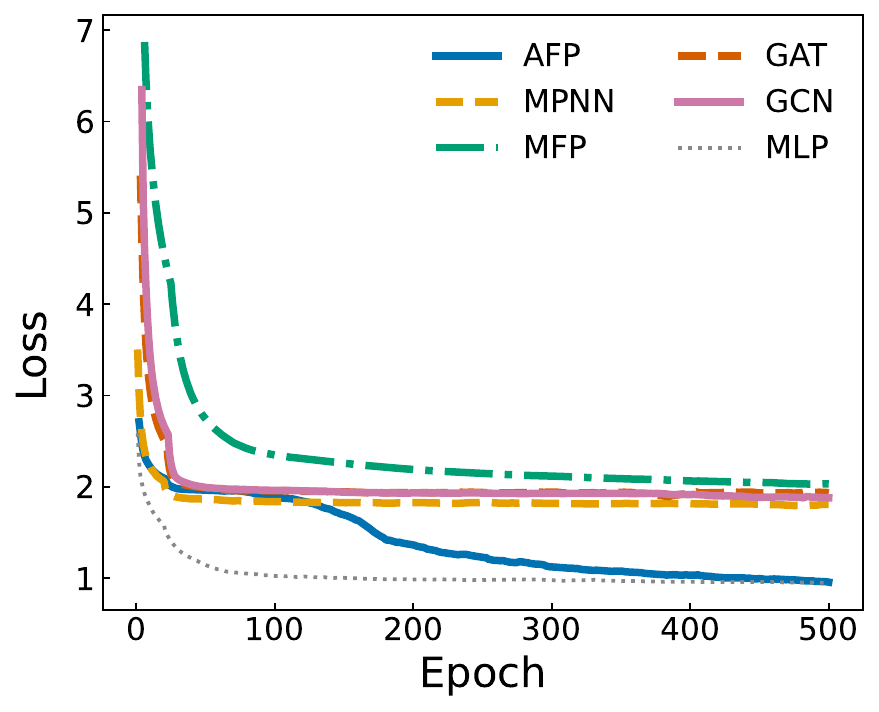}
\caption{Validation loss (EMD) curves of the five GNN models, in comparison with the MLP model trained with ECFP (dotted line).} 
\label{F2}
\end{figure}

\begin{table*}
\renewcommand{\arraystretch}{1.5}
\centering
\caption{Distance metrics compared in this work. Let $\mathbf{a}=\left(a_i\right)_{i=1}^N$ and $\mathbf{b}=\left(b_i\right)_{i=1}^N$ be two normalized spectra as two aligned histograms with the same set of bins, $D$ is their distance.}
\label{tab:gnn_models}
\begin{tabular*}{\linewidth}{@{\extracolsep{\fill}}lll@{}}
\hline
\textbf{Metric} & \textbf{Expression} & \textbf{Key idea} \\ 
\hline
EMD  & $D = \sum_{i=1}^{N} \left\vert \sum_{j\le i}\left(a_j - b_j\right)\right\vert$ & Minimum ``work'' needed to transform $\mathbf{a}$ into $\mathbf{b}$ \\
JSD  & $D = \frac{1}{2} \mathrm{KL}(\mathbf{a} \parallel \mathbf{m}) + \frac{1}{2} \mathrm{KL}(\mathbf{b} \parallel \mathbf{m}), 
\quad \mathbf{m} = \frac{\mathbf{a} + \mathbf{b}}{2}$ & Symmetrized version of Kullback-Leibler divergence \\
HD   & $D = \frac{1}{2} \sum_{i=1}^{N} \left( \sqrt{a_i} - \sqrt{b_i} \right)^2 $ & Euclidean distance between square-rooted $\mathbf{a}$ and $\mathbf{b}$ \\
TVD  & $D = \frac{1}{2} \sum_{i=1}^{N} | a_i - b_i |$ & Measures the total probability mass that differs between $\mathbf{a}$ and $\mathbf{b}$ \\
SIS  & $D = 1 - \frac{\sum a_i b_i}{\sqrt{\sum a_i^2 \sum b_i^2}}$ & Compares integrated spectral intensities over frequencies \\
\hline
\end{tabular*}
\end{table*}

Figure \ref{F2} presents the validation loss curves for the five evaluated models, confirming training convergence and enabling comparative assessment of their suitability for spectral prediction. Among these, AFP achieves the lowest loss among the GNN baselines evaluated here, although its performance remains below that of the ECFP+MLP model for several spectral regions. This is consistent with previous reports on attention-based GNNs for molecular property prediction \citep{ShabanTameh2024IRPrediction, xu2025pretrained}. This advantage stems from AFP's dual application of attention mechanisms, both during message passing and in the global readout step, which enables the model to selectively emphasize task-relevant molecular features.

The GCN, GAT, and MPNN models exhibit moderate, comparable performance levels. While GCN aggregates neighboring atoms with equal weighting and GAT employs attention to differentiate their contributions, both approaches appear insufficient for fully capturing feature importance during the final molecular representation. MPNN offers greater flexibility by explicitly incorporating both atom and bond information through its message-passing framework, making it well-suited for representing local chemical environments, yet its performance remains below that of AFP.

The MFP model, designed as a neural analog of classical fingerprints like ECFP \citep{rogers2010extended}, shows the highest validation loss, suggesting that fingerprint-style representations, whether fixed or learned, are less effective for spectral prediction tasks. Notably, a simple multilayer perceptron (MLP) trained on ECFP features outperforms all five graph-based models, as indicated by the gray curve in Figure \ref{F2}. The strong performance of the MLP baseline suggests that fixed-radius circular fingerprints retain advantages in extrapolation, possibly because they capture local aromatic environments more robustly than SMILES-derived graph embeddings.

\begin{figure}
\centering
\includegraphics[width=0.46\textwidth]{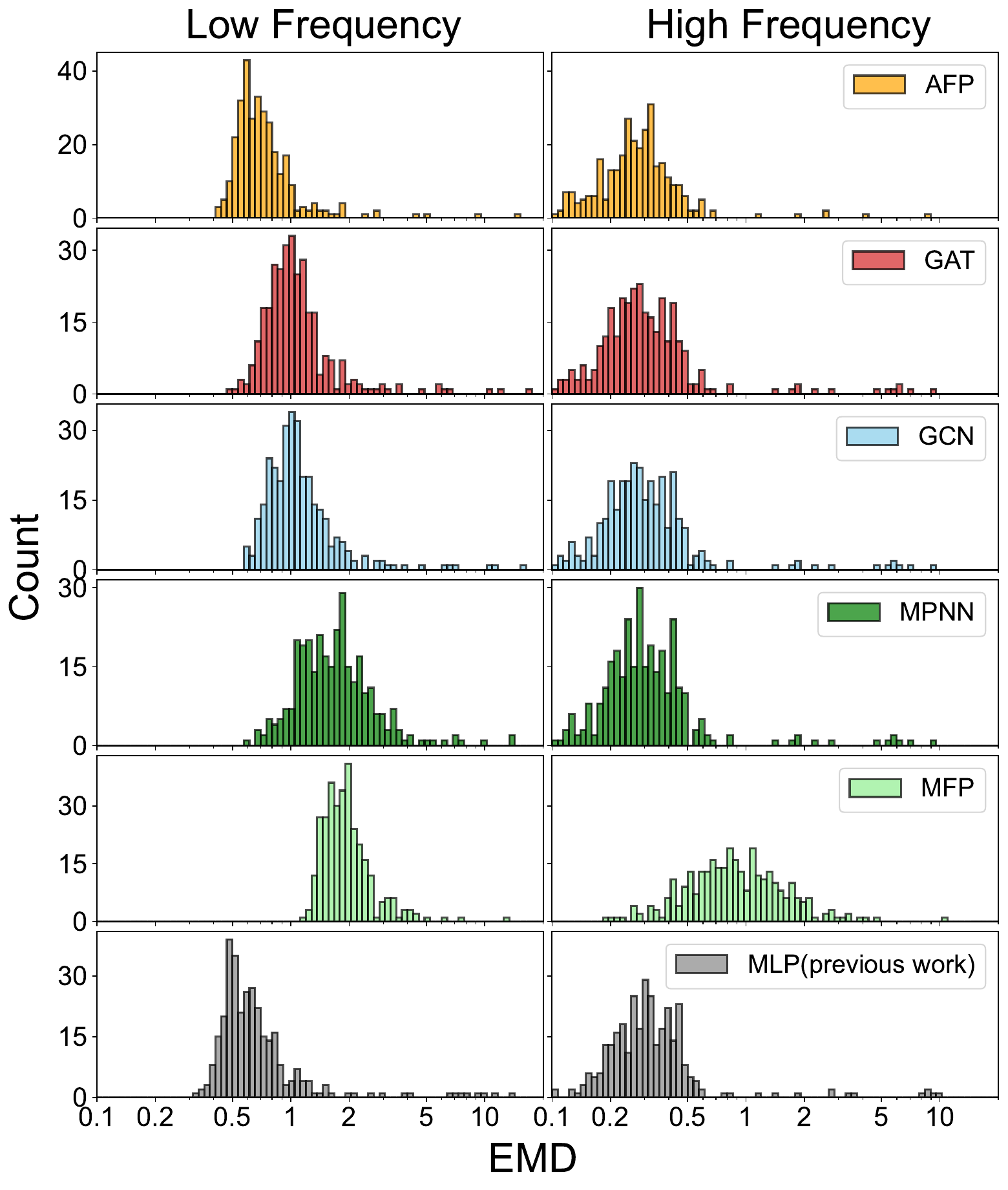}
\caption{Distribution of prediction errors for the IR spectra generated by five GNN models and the baseline MLP model, compared against high-level DFT results in the low (left) and high (right) frequency regions.}
\label{F3}
\end{figure}

Figure \ref{F3} compares the EMD distributions across models for low- and high-frequency spectral regions, revealing a distinct performance pattern. High-frequency predictions demonstrate greater accuracy, with errors concentrated at smaller EMD values (except for MFP), while low-frequency distributions show broader spreads toward larger errors. This discrepancy reflects the fundamental challenge in reproducing low-frequency vibrational modes, which are more sensitive to subtle chemical environments and long-range interactions. High-frequency modes such as \ce{C-H} and \ce{O-H} stretching are highly localized and primarily depend on specific bond strengths, making them more amenable to machine learning capture. In contrast, low-frequency modes often involve collective motions, torsions, and lattice-like vibrations that are strongly influenced by long-range couplings and anharmonic effects, presenting greater modeling challenges \citep{mcgill2021predicting}.

Within this overall trend, AFP achieves the narrowest and most centered distributions in both spectral regions, confirming its advantage in extracting relevant structural features. GCN, GAT, and MPNN demonstrate moderate performance, though their broader error distributions highlight limitations in capturing frequency-specific information. The MFP model shows the weakest performance, while the baseline MLP trained on ECFP descriptors emerges as the most reliable overall, delivering stable predictions across both frequency ranges.

The choice of an appropriate loss function is crucial for spectral prediction tasks, as effective metrics must accommodate slight frequency mismatches that general-purpose measures like Mean Absolute Error (MAE) or Root Mean Squared Error (RMSE) cannot adequately handle. While EMD has demonstrated particular suitability for this task in our previous work \citep{kovacs2020machine} and the results presented above, we further investigated four specialized distance metrics, JSD, HD, TVD, and SIS, to comprehensively evaluate their performance, as summarized in Table \ref{tab:gnn_models}.

Based on these spectral distance metrics, we constructed five distinct AFP models, each optimized using a different loss function, to predict IR spectra of PAHs via 5-fold cross-validation. For fair comparison, the distance between predicted and reference spectra was evaluated using all four remaining metrics (excluding the one used for training). Model performance was assessed based on the average rank across these metrics, as summarized in Table \ref{tab:combined_table}.

\begin{table*}[t]
    \centering
    \caption{Performance of five AFP models with the loss functions listed in Table~1, evaluated on low- and high-frequency data under four additional distance metrics.}
    \label{tab:combined_table}
    \begin{ruledtabular}
    \begin{tabular}{cccccccc}
        Freq. &  Loss & \multicolumn{5}{c}{Rank} & Average \\     
        Region & Func. & EMD & HD & JSD & SIS & TVD & Rank \\
        \colrule
        & EMD & * & 3 & 4 & 4 & 1 & 3 \\
        & HD  & 3 & * & 1 & 1 & 3 & 2 \\
   Low  & JSD & 2 & 1 & * & 2 & 2 & \textbf{1.75} \\
        & SIS & 4 & 4 & 2 & * & 4 & 3.5 \\
        & TVD & 1 & 2 & 2 & 3 & * & 2 \\
        \colrule
        & EMD & * & 2 & 3 & 2 & 2 & 2.25 \\
        & HD  & 4 & * & 2 & 3 & 1 & 2.5 \\
   High & JSD & 3 & 1 & * & 1 & 4 & 2.25 \\
        & SIS & 2 & 3 & 4 & * & 3 & 3.5 \\
        & TVD & 1 & 4 & 1 & 4 & * & 2.5 \\
    \end{tabular}
    \end{ruledtabular}
\end{table*}

Based on the results in Table \ref{tab:combined_table}, the JSD-based model achieves the best overall performance in the low-frequency region, with an average rank of 1.75, while HD- and TVD-based models show moderate performance with average ranks of 2.0. The SIS-based model consistently performs worst, with an average rank of 3.5. In the high-frequency region, all models exhibit more comparable performance, reflecting the inherent learnability of high-frequency vibrational modes. These findings indicate that JSD serves as the most robust metric for IR spectral prediction, particularly in the challenging low-frequency region where metric performance diverges most significantly. JSD is a bounded, symmetric measure of dissimilarity between two probability distributions. It compares each distribution to their average, quantifying how far apart they are. This makes JSD sensitive to both the frequency and shape of spectral features, offering a robust assessment of spectral prediction quality.

The superior performance of JSD can be attributed to the nature of low-frequency spectral features. Low-frequency vibrations typically involve broad, overlapping modes that are better captured by distribution-based similarity measures like JSD, which emphasize overall spectral shape rather than precise peak alignment. In contrast, alignment-sensitive metrics such as HD, TVD, and SIS are more susceptible to small frequency shifts and broad peak structures, reducing their effectiveness in this spectral region. This explains why JSD excels particularly in low-frequency prediction, while performance differences between metrics become less pronounced at higher frequencies.

\begin{figure}
\centering
\includegraphics[width=0.46\textwidth]{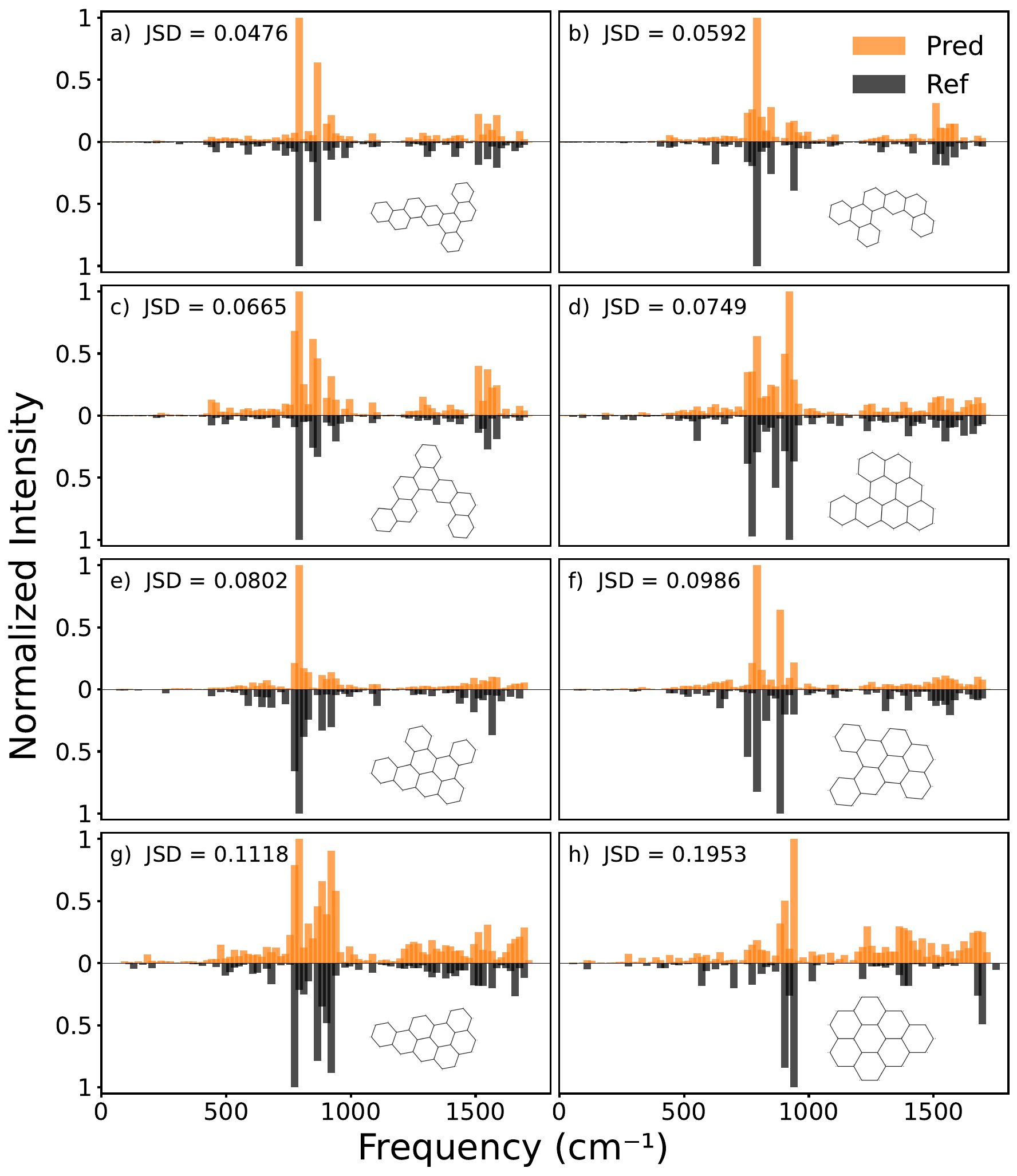}
\caption{Comparison of predicted (upper) and reference (lower) IR spectra for eight PAHs selected across the JSD error distribution, representing increasing prediction error from best (a) to poorest (h). All spectra are normalized to their maximum intensity.}
\label{F4}
\end{figure}

To illustrate the range of prediction quality, Figure~\ref{F4} compares predicted and reference low-frequency spectra for eight representative PAHs, ordered from most to least accurate. These predictions were generated by the AFP model trained with the JSD loss function, with panels (a)-(h) sampled across the JSD error distribution. In the most accurate cases (panels a-c), the model successfully captures both peak positions and intensity distributions, demonstrating its capability to reproduce complex low-frequency spectral features.

\begin{figure}
\centering
\includegraphics[width=0.45\textwidth]{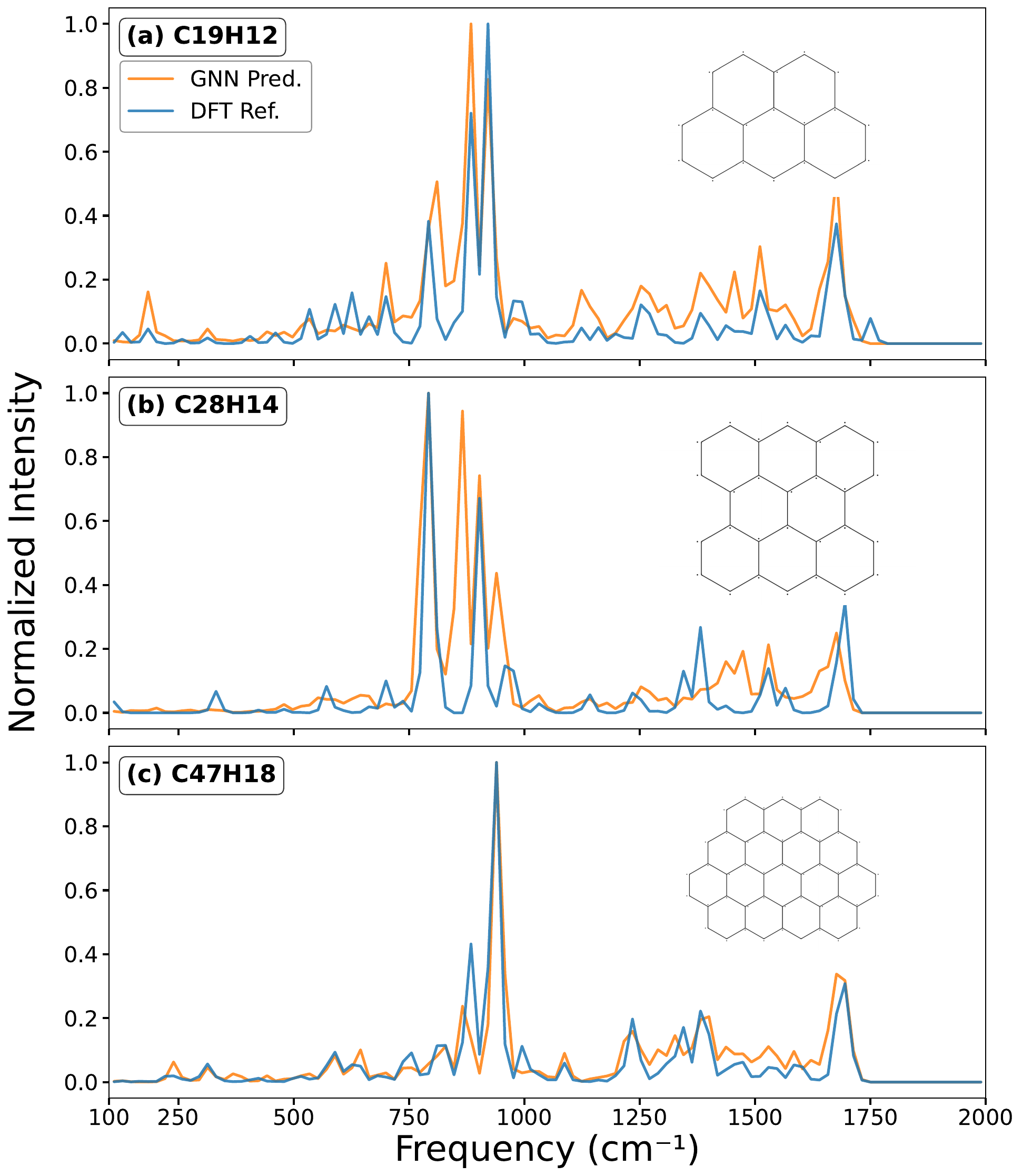}
\caption{Comparison of GNN-predicted and DFT-computed infrared spectra for three representative pericondensed molecules. The spectra are broadened with a Gaussian line profile (FWHM = \SI{10}{\per\centi\meter}) \citep{Yang2020Superhydrogenated}.}
\label{F5}
\end{figure}

As prediction quality declines from panels (d) to (h), discrepancies between predicted and reference spectra become more apparent, manifesting as underestimated or overestimated peak intensities, occasional frequency shifts, and missing spectral features. Despite these inaccuracies, even the poorest predictions retain the overall spectral shape and capture major bands, demonstrating the model's fundamental capability to learn meaningful structure-spectrum relationships. 

Figure \ref{F5} compares the IR spectra predicted by the GNN model against the reference DFT spectra for three representative pericondensed PAH molecules. This comparison further validates the effectiveness of the GNN framework, particularly for molecular structures that differ from the catacondensed PAHs shown in Figure 4. The results demonstrate strong agreement between the GNN predictions and DFT calculations, especially in key spectral regions. While minor discrepancies appear at specific wavenumbers, the overall spectral features, including peak positions and intensities, are consistently well-represented.

\begin{figure}
\centering
\includegraphics[width=0.95\linewidth]{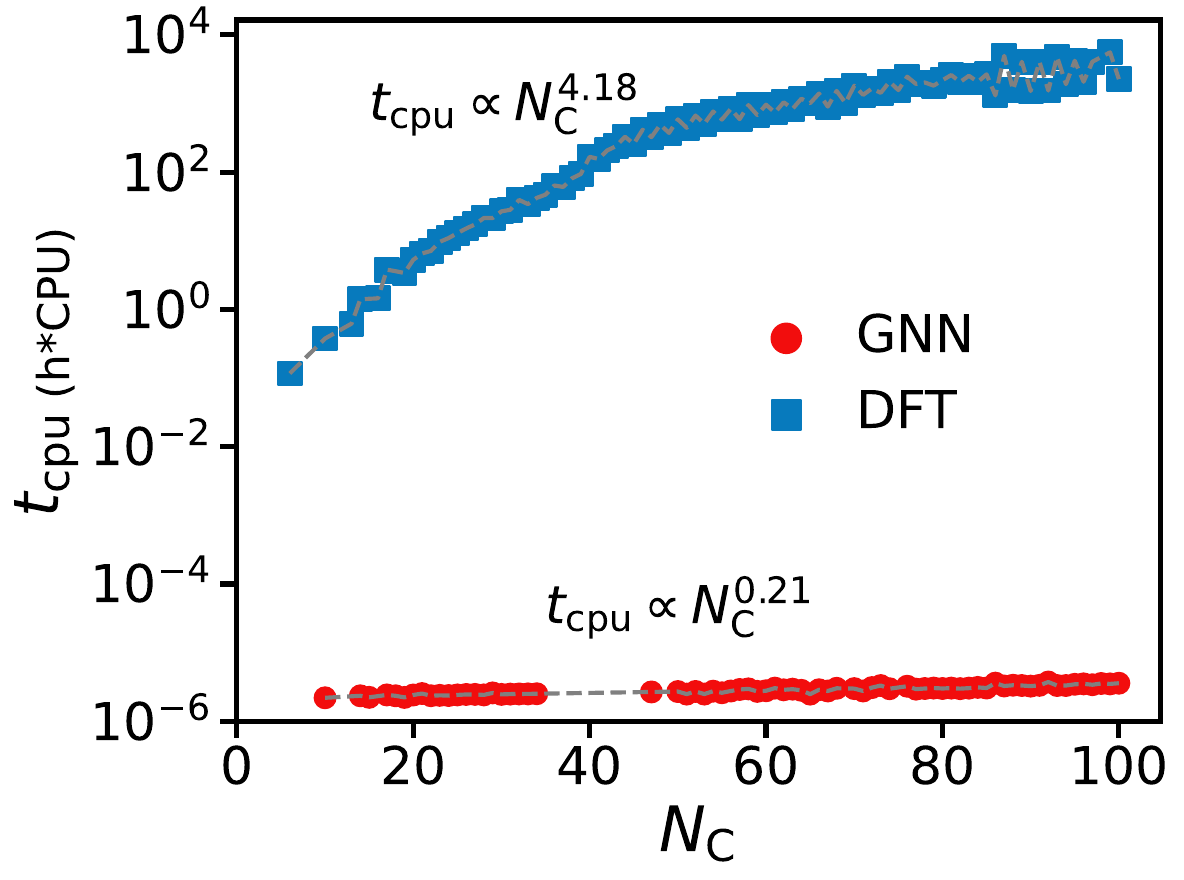}
\caption{Computational time for calculating harmonic IR spectra versus PAH size ($N_\mathrm{C}$). Results compare the AFP model (sequential execution on Intel I5-13500HX CPU) with DFT calculations at B3LYP/4-31G level (parallelized across 40 Intel Xeon E5-2680 CPU cores).}
\label{F7}
\end{figure}

Despite delivering comparable accuracy, our GNN model demonstrates dramatically superior computational efficiency relative to conventional quantum chemical calculations. As shown in Figure~\ref{F7}, the AFP model computes IR spectra two to five orders of magnitude faster than DFT, achieving performance comparable to machine-learning molecular dynamics (MLMD) approaches \citep{laurens2021infrared}. The GNN inference time exhibits only weak dependence on molecular size, scaling as $t_\mathrm{GNN} \propto N_\mathrm{C}^{0.21}$, indicative of near-linear computational complexity. In contrast, DFT computation time increases steeply as $t_\mathrm{DFT} \propto N_\mathrm{C}^{4.18}$, reflecting the substantial cost of electronic structure calculations. For PAHs exceeding approximately 40 C atoms, the GNN model achieves speedups greater than 10,000-fold compared to DFT. This dramatic efficiency gain underscores the practical potential of GNN frameworks for large-scale spectral generation, aligning with recent advances in machine-learning accelerated molecular simulation \citep{mai2025computing}.

\begin{figure}
\centering
\includegraphics[width=0.9\linewidth]{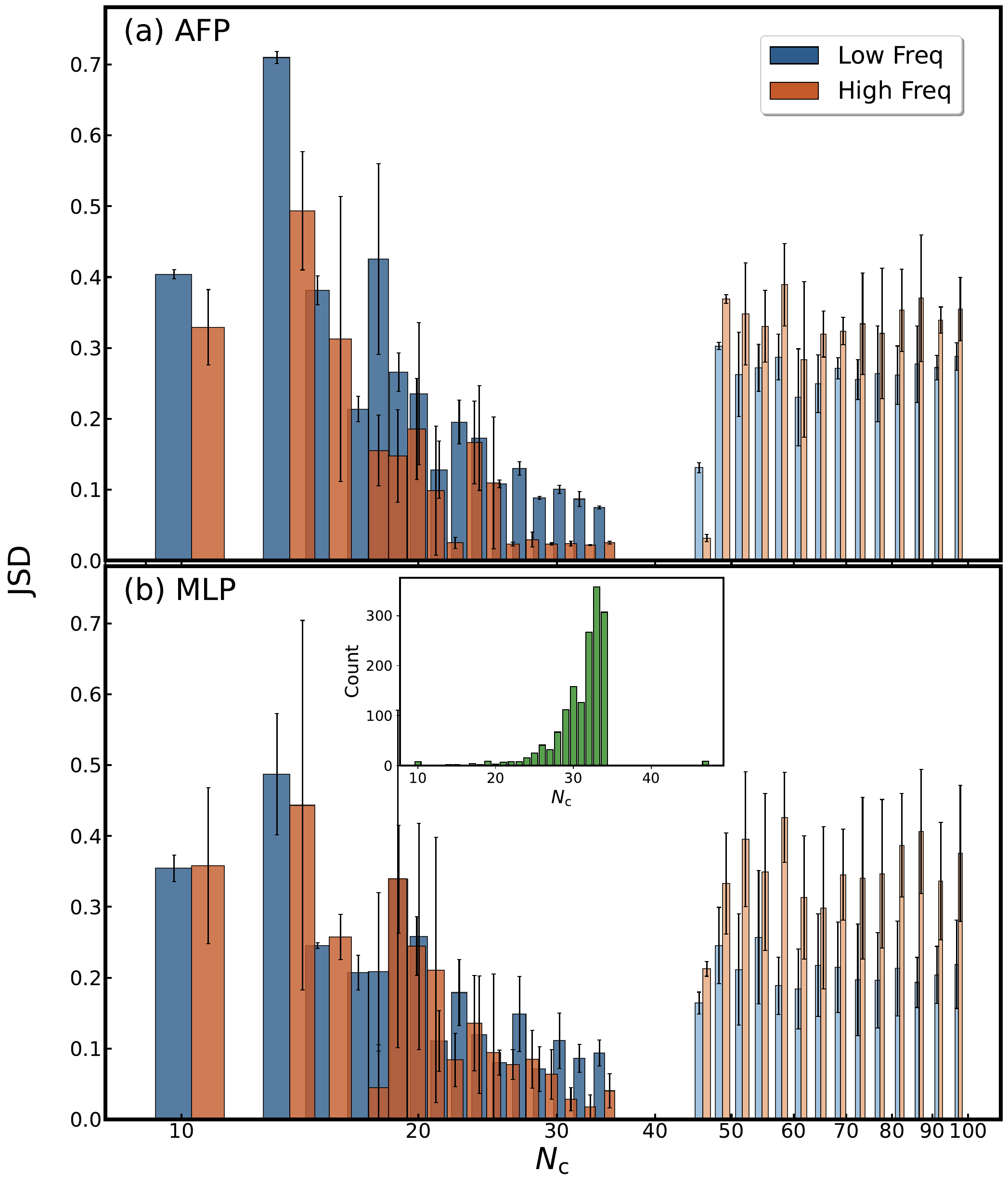}
\caption{Average prediction error versus molecular size ($N_\mathrm{C}$, log scale) for (a) the AFP model and (b) the previous MLP model. Blue and orange curves represent low- and high-frequency regions, respectively. Inset: molecular size distribution in the training dataset.}
\label{F6}
\end{figure}

A critical consideration for astronomical applications is model performance on large PAH molecules ($N_\mathrm{C}>50$), which are proposed as dominant carriers of AIBs \citep{sellgren1984near, Allamandola1989} yet remain severely underrepresented in current databases. To evaluate extrapolation capability, we applied our best-performing AFP+JSD model to large PAHs excluded from training, comparing its performance against our previously developed MLP model. Figure \ref{F6} displays the average spectral error as a function of molecular size ($N_\mathrm{C}$) for both models, with panel (a) showing AFP results and panel (b) the MLP baseline. The inset illustrates the training set's molecular size distribution, highlighting the scarcity of large PAHs.
  
Both models exhibit consistently smaller errors in the high-frequency region compared to the low-frequency domain, aligning with trends observed in Figure \ref{F3}. This performance gap stems from the localized nature of high-frequency modes (e.g., C–H stretching), which are inherently more learnable than the complex, delocalized low-frequency vibrations. The AFP model maintains low, stable errors for intermediate-sized PAHs ($N_\mathrm{C}$ ranging from 21 to 34), where training data are most abundant, highlighting its data-driven characteristics.

When evaluating generalization to larger molecules ($N_\mathrm{C} > 40$), the AFP model shows increasing errors in both spectral regions, indicating limited extrapolation capability for large PAHs. This indicates that the model should not be applied to astronomical PAHs outside the training domain. In contrast, the MLP model demonstrates a more gradual error increase with molecular size, suggesting slightly better robustness for low-frequency extrapolation. This pattern aligns with established limitations of data-driven models when applied beyond their training domain \citep{meng2021machine, beglaryan2025toward}. 

\section{Conclusions}
In conclusion, this work shows that GNNs, particularly the AFP architecture, provide a computationally efficient and chemically intuitive framework for predicting molecular IR spectra in astronomical applications. A key finding is the decisive role of the loss function: the JSD outperforms traditional metrics by better capturing broad, overlapping low-frequency bands, emphasizing that spectral prediction depends on overall distribution rather than exact point-wise alignment. Notably, the GNN predicts IR spectra orders of magnitude faster than conventional quantum chemical calculations, with near-linear scaling in molecular size, making it potentially useful for generating approximate spectra of small-medium PAHs, while further work is required before application to astronomical PAH populations.

Despite their strong performance, the GNN models remain data-driven and show reduced accuracy for PAHs larger than approximately 40 C atoms, reflecting the limited coverage of current datasets. Future improvements may come from incorporating physics-based priors or hybrid quantum-ML schemes. Beyond methodology, this framework may contribute to future IR studies once additional training data and physics-based constraints are incorporated. Another limitation of this work is that the model relies solely on SMILES-based topology without incorporating 3D geometric or physics-informed descriptors. Consequently, its predictive power for structurally complex PAHs remains limited. Future models incorporating geometry or equivariant message passing may improve extrapolation.

%

\end{document}